\documentstyle[12pt]{article}
\makeatletter
\@addtoreset{equation}{section}

\makeatother

\begin{document}

\begin{flushright}
Feb 2001

OU-HET 380
\end{flushright}

\begin{center}

\vspace{5cm}

{\Large Tachyon Condensation and} 

\vspace{3mm}

{\Large Spectrum of Strings on D-branes}

\vspace{2cm}

Takao Suyama \footnote{e-mail address : suyama@funpth.phys.sci.osaka-u.ac.jp}

\vspace{1cm}

{\it Department of Physics, Graduate School of Science, Osaka University, }

{\it Toyonaka, Osaka, 560-0043, Japan}

\vspace{4cm}

{\bf Abstract} 

\end{center}

We investigate spectrum of open strings on D-branes after tachyon condensation in bosonic 
string theory. 
We calculate 1-loop partition function of the string and show that its limiting forms 
coincide with partition functions of open strings with different boundary conditions. 

\newpage

\vspace{1cm}

\section{Introduction}

\vspace{5mm}

The study on unstable D-brane systems in recent years has taught us many new aspects of 
string theory \cite{TC}. 
It is believed that the unstable D-brane systems decay into the perturbative vacuum of 
closed strings or lower dimensional stable D-branes. 
The occurrence of such phenomena has been shown by some analyses \cite{TC}\cite{SFT}. 
These phenomena are understood in terms of tachyon condensation, i.e. condensation of open 
string tachyons. 
In other words, these researches have clarified the mechanism of the stabilization of the 
tachyons. 
The occurance of the tachyon condensation had also been investigated in \cite{history}. 

Some of the unstable D-branes can be described by a class of worldsheet theory \cite{HKM} 
whose action contains a boundary term representing the effect of the tachyon as well as the 
bulk action. 
In general, the boundary term can be regarded as a relevant parturbation, so the 
condensation of the tachyon can be studied by analyzing the renormalization group flow 
induced by the boundary term. 
The simplest choice of the boundary term is 
\begin{equation}
u\int_{\partial\Sigma}d\xi\ X(\xi)^2.
  \label{boundary}
\end{equation}
Since the worldsheet theory with this term remains a free theory, some exact results can be 
extracted from this theory. 
One remarkable example is the result obtained by boundary string field theory \cite{bSFT} 
which enables one to derive an effective action of spacetime fields from the disk partition 
function of the above theory. 
The exact value of the D-brane tension is also obtained. 
For this purpose, it is suitable to compactify the theory on $S^1$ and this leads one to 
the study of an integrable system \cite{FI}. 

To identify the decay remnants as D-branes, it should also be checked that the physical 
degrees of freedom on the remnants coincide with that of the open string on D-branes. 
This has been considered in \cite{toy} by using toy models which capture some properties 
of the unstable brane system. 

In this paper we argue this problem in terms of the worldsheet theory perturbed by the 
boundary term introduced above. 
The spectrum of an open string can be obtained by calculating 1-loop (annulus) partition 
function. 
We perform the calculation of the partition function including the boundary term 
(\ref{boundary}) and show that this connects continuously the partition functions of open 
strings with different boundary conditions. 
Similar analyses on the effect of boundary terms have been done \cite{bCFT}. 
Thrust of this paper is to show a connection between the Neumann-Neumann (NN) 
strings (or the Dirichlet-Dirichlet (DD) strings) and the Neumann-Dirichlet (ND) strings 
as well as between the NN strings and the DD strings. 
They are shown by explicitly deriving their partition functions as limiting forms of the 
same function. 

This paper is organized as follows. 
In section \ref{classical} the effect of the boundary term (\ref{boundary}) is shown in view 
of classical solutions. 
To obtain the 1-loop partition function, we calculate in section \ref{greens} the Green's 
function on annulus with appropriate boundary conditions. 
This is used to obtain the partition function in section \ref{partition}. 
A possible extension and a suggestion of this result to the structure of string theory are 
discussed in section \ref{discuss}.

\vspace{1cm}

\section{Classical solution} \label{classical}

\vspace{5mm}

The phenomena of tachyon condensation in bosonic string theory can be described by the 
following very simple action \cite{HKM}. 
\begin{equation}
S=\frac1{4\pi\alpha'}\left[ \int_{\Sigma}d\tau d\sigma\ \partial_aX_\mu\partial_aX^\mu
     +\int_{\partial\Sigma}d\xi\ \sum_{\nu=p}^{25}u_\nu (X^\nu)^2 \right]
  \label{action}
\end{equation}
$\xi$ parametrizes the boundary $\partial\Sigma$ of the worldsheet $\Sigma$. 
The couplings $u_\nu$ increase along the renormalization group flow, and in the IR limit 
($u_\nu\to+\infty$) the second term of the action (\ref{action}) imposes $X^\nu =0$. 
So the Neumann boundary condition in the UV limit turns into the Dirichlet boundary 
condition in the IR limit. 
This indicates the decay of the original D25-brane into a Dp-brane at $x^\nu=0$. 

\vspace{2mm}

This can also be seen by solving the equation of motion, taking into account the boundary 
term. 
To see the effect of the boundary term, we discuss the following situation. 
Let us consider two Dp-branes and an open string stretched between them. 
If one of the Dp-branes decays into a D(p$-$1)-brane, one of the coordinates of the open 
string will obey the Neumann-Dirichlet condition (i.e. the Neumann condition at one end and 
the Dirichlet condition at the other end) after the decay. 

We will focus on the coordinate $X$ along which the Dp-brane shrinks. 
The solution of the equation of motion is as follows, 
\begin{equation}
X(z,\bar{z}) = \sum_{\lambda\in\Lambda} 
                  (a_{\lambda}z^{-\lambda}+\tilde{a}_\lambda \bar{z}^{-\lambda})
  \label{X}
\end{equation}
where $z=e^{\tau+i\sigma}$ and $\tau,\sigma$ are the worldsheet coordinates. 
The scalar $X$ must satisfy the boundary conditions
\begin{equation}
\left\{
  \begin{array}{ll}
    \partial_\sigma X=0, & (\sigma=0) \\ \partial_\sigma X+uX=0. & (\sigma=\pi)
  \end{array}
\right.
\end{equation}
These conditions are satisfied if all $\lambda$'s are solutions of the following equation. 
\begin{equation}
\lambda \tan(\lambda\pi)=-u
  \label{moding}
\end{equation}
This equation can be easily solved when $u=0,+\infty$ \ i.e. UV and IR limit. 
\begin{equation}
\lambda\in\left\{
  \begin{array}{ll}
    {\bf Z} & (u=0) \\ {\bf Z}+\frac12 & (u=\infty)
  \end{array}
\right.
\end{equation}
So one can see that the NN string (integral moding) turns continuously into the ND string 
(half-integral moding) as $u$ increases. 

\vspace{2mm}

In principle, one can calculate, for example, 1-loop partition function with general $u$ 
by quantizing the open string coordinate (\ref{X}). 
However, solving the eq.(\ref{moding}) is difficult in general $u$. 
Moreover the $\lambda$'s are not at regular intervals, so the calculations will be very hard. 
In the following sections, we will take another way to calculate the 1-loop partition 
function.

\vspace{1cm}

\section{Green's function on annulus} \label{greens}

\vspace{5mm}

In this section, we calculate Green's function of the scalar field $X$ on 
annulus $A$ \cite{annulus}. 
\begin{equation}
A = \left\{ z\in {\bf C} \ | \ r\le|z|\le1 \right\}
\end{equation}
$r$ is related to the modulus of the annulus. 

Again we consider an open string stretched between two Dp-branes. 
Then each end of the string interacts with tachyon field on each Dp-brane. 
Therefore the action of the open string has, in general, different boundary term for each 
boundary of the annulus $A$. 
This corresponds to imposing different boundary condition on the Green's function for 
each boundary.  

The Green's function $G(z,w)$ we would like to find is the solution of the equation 
\begin{equation}
-\frac1{\pi\alpha'}\partial_z\partial_{\bar{z}}G(z,w)=\delta^2(z,w),
  \label{Poissoneq}
\end{equation}
satisfying the following boundary conditions 
\begin{eqnarray}
\Bigl[z\partial_z+\bar{z}\partial_{\bar{z}}+u\Bigr]G(z,w)|_{|z|=1} &=& 0, 
  \label{bdy1} \\
\Bigl[-r^{-1}(z\partial_z+\bar{z}\partial_{\bar{z}})+v\Bigr]G(z,w)|_{|z|=r} &=& 0.
  \label{bdy2}
\end{eqnarray}
$u,v=0$ corresponds to the NN condition, $u,v\to+\infty$ to the DD condition and 
$u=0,v\to+\infty$ (and $u\to+\infty,v=0$) to the ND condition. 
In addition, $G(z,w)$ should have the following properties. 
\begin{eqnarray}
G(z,w) &=& G(w,z) \nonumber \\
G(e^{i\varphi}z,e^{i\varphi}w) &=& G(z,w) \hspace{1cm} \mbox{(rotation around the origin)}
\end{eqnarray}

We start with the following ansatz. 
\begin{eqnarray}
G(z,w) &=& -\frac{\alpha'}2\log|z-w|^2+a\log|z|^2\log|w|^2+b(\log|z|^2+\log|w|^2)+c 
           \nonumber \\
& & +\sum_{k=1}^\infty d_k\left\{ \left(\frac zw\right)^k+\left(\frac wz\right)^k\right\}
    +\sum_{k=1}^\infty \tilde{d}_k\left\{ \left(\frac{\bar{z}}{\bar{w}}\right)^k
                                         +\left(\frac{\bar{w}}{\bar{z}}\right)^k\right\}
    \nonumber \\
& & +\sum_{k=1}^\infty e_k\left\{(z\bar{w})^k+(\bar{z}w)^k\right\}
    +\sum_{k=1}^\infty \tilde{e}_k\left\{(z\bar{w})^{-k}+(\bar{z}w)^{-k}\right\}
\end{eqnarray}
This is, of course, a solution of the eq.(\ref{Poissoneq}). 
The boundary conditions (\ref{bdy1})(\ref{bdy2}) are now reduced to linear equations of the 
coefficients. 

The condition (\ref{bdy1}) implies, for each $k$, 
\begin{eqnarray}
2a+ub &=& 0, \nonumber \\
2b+uc &=& \alpha', \nonumber \\
(k+u)d_k+(-k+u)\tilde{e}_k &=& 0, \\
(-k+u)d_k+(k+u)e_k &=& \frac{\alpha'}2\left(1-\frac uk\right), \nonumber \\
(k+u)\tilde{d}_k+(-k+u)\tilde{e}_k &=& 0, \nonumber \\
(-k+u)\tilde{d}_k+(k+u)e_k &=& \frac{\alpha'}2\left(1-\frac uk\right). \nonumber
\end{eqnarray}

The condition (\ref{bdy2}) means 
\begin{eqnarray}
\frac{2a}r-2av\log r-bv &=& -\frac{\alpha'}2v, \nonumber \\
\frac{2b}r-2bv\log r-cv &=& 0, \nonumber \\
\left(\frac kr-v\right)d_k-\left(\frac kr+v\right)r^{-2k}\tilde{e}_k 
&=& -\frac{\alpha'}2\left(\frac1r-\frac vk\right), \\
-\left(\frac kr+v\right)d_k+\left(\frac kr-v\right)r^{2k}e_k &=& 0, \nonumber \\
\left(\frac kr-v\right)\tilde{d}_k-\left(\frac kr+v\right)r^{-2k}\tilde{e}_k
&=& -\frac{\alpha'}2\left(\frac1r-\frac vk\right), \nonumber \\
-\left(\frac kr+v\right)\tilde{d}_k+\left(\frac kr-v\right)r^{2k}e_k &=& 0. \nonumber 
\end{eqnarray}

These equations have the following solution, 
\begin{eqnarray}
a &=& \frac{\alpha'}4uv\left(-\frac ur+uv\log r-v\right)^{-1}, \nonumber \\
b &=& \frac{\alpha'}2v\left(\frac ur-uv\log r+v\right)^{-1}, \nonumber \\
c &=& -\frac{\alpha'}2\left(\frac2r-2v\log r\right)^{-1}
       \left(-\frac ur+uv\log r-v\right)^{-1}, \\
d_k &=& \tilde{d}_k = (k-u)\left(\frac kr-v\right)r^{2k}f_k, \nonumber \\
e_k &=& (k-u)\left(\frac kr+v\right)f_k, \nonumber \\
\tilde{e}_k &=& (k+u)\left(\frac kr-v\right)r^{2k}f_k, \nonumber
\end{eqnarray}
where 
\begin{equation}
f_k = -\frac{\alpha'}{2k}\left\{(k-u)\left(\frac kr-v\right)r^{2k}
-(k+u)\left(\frac kr+v\right)\right\}^{-1}. 
\end{equation}

\vspace{1cm}

\section{1-loop partition function} \label{partition}

\vspace{5mm}

From the Green's function obtained in the previous section, one can calculate the 
expectation value of the energy-momentum tensor. 
From the following OPE
\begin{equation}
\partial X(z)\partial X(w) = -\frac{\alpha'}2\frac1{(z-w)^2}-\alpha'T_{ww}(w)
                             +O(z-w)
\end{equation}
where 
\begin{equation}
T_{zz} = -\frac1{\alpha'}\partial X\partial X,
\end{equation}
one can obtain 
\begin{eqnarray}
\langle T_{zz}(z) \rangle &=& -\frac1{\alpha'}\lim_{w\to z}
        \left[\partial_z\partial_wG(z,w)+\frac{\alpha'}2\frac1{(z-w)^2}\right] \nonumber \\
&=& -\frac1{\alpha'}\frac1{z^2}\left(a-\sum_{k=1}^\infty 2k^2d_k\right).
\end{eqnarray}
Similar result can be obtained for $\langle \bar{T}_{\bar{z}\bar{z}}(\bar{z})\rangle$. 

It is convenient to map the annulus $A$ into rectangle $R$. 
\begin{equation}
R = \{ \rho\in {\bf C}\ |\ -\pi\le\Re\rho\le 0,\ 0\le\Im\rho\le 2\pi t\}
\end{equation}
$z$ and $\rho$ are related as $z=\exp(\rho/t)$, where $t=-\pi/\log r$. 

The expectation value of $T_{\rho\rho}(\rho)$ on $R$ is 
\begin{eqnarray}
\langle T_{\rho\rho}(\rho) \rangle
&=& \left(\frac{dz}{d\rho}\right)^2\langle T_{zz}(z)\rangle +\frac c{12}\{z,\rho\} 
                       \nonumber \\
&=& -\frac1{t^2}\left[\frac1{\alpha'}\left(a-\sum_{k=1}^\infty 2k^2d_k\right)+\frac1{24}
     \right].
\end{eqnarray}                  

\vspace{2mm}

The change of the modulus of the worldsheet corresponds to a change of the metric. 
Therefore the variation of the 1-loop partition function $Z$ with respect to the modulus is 
proportional to the expectation value of the energy-momentum tensor. 
The precise relation we need is as follows. 
\begin{equation}
\delta\log Z = -\frac1{2\pi}\int d^2\rho\ 
  \left(\delta g_{\rho\rho}\langle\bar{T}_{\bar{\rho}\bar{\rho}}(\bar{\rho})\rangle
       +\delta g_{\bar{\rho}\bar{\rho}}\langle T_{\rho\rho}(\rho)\rangle\right)
\end{equation}
If the length $t$ of the rectangle $R$ changes by $\delta t$, the corresponding change of 
the metric is 
\begin{equation}
\delta g_{\rho\rho}=\delta g_{\bar{\rho}\bar{\rho}}=-\frac1{2t}\delta t,
\end{equation}
up to a rescaling of the metric. 

Now we obtain a differential equation for the 1-loop partition function
\begin{equation}
\frac d{dt}\log Z(t) = 
  -\frac{2\pi}{t^2}\left[\frac1{\alpha'}\left(a-\sum_{k=1}^\infty 2k^2d_k\right)+\frac1{24}
  \right].
    \label{difeq}
\end{equation}

Note that since the conformal invariance is broken at the boundary for general $u$ and $v$, 
the argument above might not be valid in general. 
Nevertheless, the equation (\ref{difeq}) is meaningful at least at the conformal limits of 
$u,v$ which we would like to discuss. 

\vspace{2mm}

Let us see whether the solution of the above equation reproduces the well-known form of the 
partition functions for (i) NN string, (ii) DD string and (iii) ND string. 

\vspace{5mm}

(i) NN string

This case corresponds to taking $u,v=0$. 
In this limit, 
\begin{equation}
a=0, \hspace{1cm} d_k=-\frac{\alpha'}{2k}\frac{r^{2k}}{r^{2k}-1}.
\end{equation}
Note that the singularity in the Green's function does not appear in the eq.(\ref{difeq}). 
The partition function is 
\begin{eqnarray}
\log Z(t) &=& -\int dt\ \frac1{t^2}\left(\sum_{k=1}^\infty 
          \frac{2\pi ke^{-2\pi k/t}}{e^{-2\pi k/t}-1}+\frac{2\pi}{24}\right)
          \nonumber \\
     &=& -\sum_{k=1}^\infty\log\left(1-e^{-2\pi k/t}\right)+\frac{2\pi}{24t}+\mbox{const.}
          \nonumber \\
     &=& -\log \eta\left(\frac it\right) +\mbox{const.}
\end{eqnarray}
Thus 
\begin{eqnarray}
Z(t) &\propto& \eta\left(\frac it\right) \nonumber \\
     &=& \frac1{\sqrt{t}}\frac1{\eta(it)}. 
\end{eqnarray}
This is the precise form including the zeromode contribution. 

\vspace{5mm}

(ii) DD string

This case corresponds to the limit $u,v\to +\infty$. 
Then 
\begin{equation}
a=\frac{\alpha'}4\frac1{\log r}, \hspace{1cm} 
d_k=-\frac{\alpha'}{2k}\frac{r^{2k}}{r^{2k}-1}.
\end{equation}
In this case, 
\begin{eqnarray}
\log Z(t) &=& -\int dt\ \frac1{t^2}\left(-\frac t2+\sum_{k=1}^\infty
               \frac{2\pi ke^{-2\pi k/t}}{e^{-2\pi k/t}-1}+\frac{2\pi}{24}\right)
               \nonumber \\
          &=& -\log\left(\frac1{\sqrt{t}}\eta\left(\frac it\right)\right).
\end{eqnarray}
Thus the correct result
\begin{equation}
Z(t) \propto \frac1{\eta(it)}
\end{equation}
is obtained. 

\vspace{5mm}

(iii) ND string

The remaining case can be realized by taking $u=0,v\to+\infty$ (the limit $u\to\infty,
v=0$ gives the same result). 
In this case, 
\begin{equation}
a=0, \hspace{1cm} d_k=-\frac{\alpha'}{2k}\frac{r^{2k}}{r^{2k}+1}.
\end{equation}
From these data, one obtains 
\begin{eqnarray}
Z(t) &\propto& e^{2\pi/24t}\prod_{k=1}^\infty\left(1+e^{-2\pi k/t}\right)^{-1} \nonumber \\
     &=& \frac{\eta(i/t)}{\eta(2i/t)} \nonumber \\
     &\propto& \frac{\eta(it)}{\eta(it/2)} \nonumber \\
     &=& q^{1/48}\prod_{k=1}^\infty\frac1{1-q^{k-1/2}}. \hspace{1cm} (q=e^{-2\pi t})
\end{eqnarray}

\vspace{1cm}

\section{Discussions} \label{discuss}

\vspace{5mm}

The calculations we have performed will also be able to apply to superstring theory. 
Consider an open string on, say, D9-brane in IIA theory. 
The spectrum of the open string is the one without GSO-projection \cite{Sen}. 
So the 1-loop partition function of the worldsheet fermions in the NS-sector will be 
\begin{equation}
Z \propto \mbox{Tr}_{NS}\ q^{L_0}.
  \label{unGSO}
\end{equation}
If the D9-brane decays into a D8-brane, the GSO-projection is imposed on the spectrum and 
the partition function in the NS-sector becomes
\begin{equation}
Z \propto \frac12\mbox{Tr}_{NS}\ q^{L_0}+\frac12\mbox{Tr}_{NS}(-1)^Fq^{L_0}.
  \label{GSO}
\end{equation}

The boundary term corresponding to this decay in the case of superstring is \cite{superbdy}
\cite{sBSFT}
\begin{equation}
u\int_{\partial\Sigma}d\xi\ \left[X^2+\psi\frac1{\partial_\xi}\psi\right].
  \label{sbdy}
\end{equation}
These terms depend only on one worldsheet fermion. 
Therefore they do not affect the other fermions. 
In the case of bosonic string, the 1-loop partition function is just a product of scalar 
partition functions. 
So the decay of D-branes can be described only by the scalars corresponding to the 
directions along which the D-branes shrink. 
On the other hand, the partition function of the fermions will not have such product form 
in general, as seen above. 
So it seems difficult that the partition function with the boundary term (\ref{sbdy}) 
connects (\ref{unGSO}) with (\ref{GSO}) continuously. 
It is interesting to see how the GSO-projection is imposed automatically in the IR limit. 

The decay of the D-branes to lower dimensional BPS branes in superstring is also interesting 
for the following reason. 
After the tachyon condensation, half of the physical degrees of freedom on the original 
unstable brane disappears due to the GSO-projection. 
This resembles the phenomenon which occurs in the decay of D-branes into the vacuum, which 
is rather difficult to analyze. 
In this case all of the open string degrees of freedom is assumed to disappear \cite{vac}. 
So discussing the former decay may be helpful for the study of the latter decay and the 
physics around the minimum of the tachyon potential. 

The calculation of the partition functions on annulus would be suitable for the study of the 
brane-antibrane systems
\footnote[1]{We thank G.Mandal for valuable discussions on this direction.}. 
The string which feels the existence of both D-brane and $\bar{\mbox{D}}$-brane is the one 
stretched between them. 
So the fundamental quantities for that system would be annulus amplitudes. 
It will be natural to expect that an annulus partition function with boundary terms 
corresponding to tachyon and gauge fields provides an effective action of the spacetime 
fields if one can relate the partition function to the spacetime action, as has been done 
for disk partition functions in boundary string field theory \cite{sBSFT}. 

\vspace{2mm}

It has been shown \cite{bSFT}\cite{FI} that the tree-level calculation provides the correct 
value of the D-brane tension. 
The form of the worldvolume is well described by the classical solutions of the spacetime 
effective field theory of open string states \cite{Suyama}. 
And we have shown in this paper that the correct spectrum on the D-branes can also be 
obtained from the 1-loop calculation. 
These results seem to indicate that, at least in the bosonic case, whole information on 
D-branes is ``built-in'' in the open string sector. 
The appropriate open strings would appear by choosing the tachyonic background, rather than 
choosing the boundary conditions of open strings by hand. 
If the same is true for superstring, it might be that the inclusion of D-branes into the 
theory of superstrings is equivalent to the inclusion of open superstrings without 
GSO-projection \cite{Yoneya}. 
We hope that such a viewpoint will be a guide to formulate the string theory. 

\vspace{1cm}

{\Large {\bf Acknowledgements}}

\vspace{5mm}

I would like to thank H. Itoyama, G. Mandal and K. Murakami for valuable discussions. 
This work is supported in part by JSPS Reseach Fellowships.

\end{document}